\documentclass[a4paper,amsmath,amssymb,nofootinbib,showpacs,notitlepage,showkeys,prd,twocolumn]{revtex4-1}

\usepackage[latin1]{inputenc}
\usepackage{graphicx}
\usepackage{hyperref}

\newcommand{\de}{\delta}
\newcommand{\eref}[1]{Eq. (\ref{#1})}
\newcommand{\fref}[1]{Fig. \ref{#1}}
\newcommand{\nnnl}{\nonumber\\}	
\newcommand{\LL}{{\Lambda^2}}
\newcommand{\LQ}{{\Lambda_\mathrm{QCD}^2}}

\allowdisplaybreaks 

\begin{document}

\title{On gluon and ghost propagators in linear covariant gauges}

\author{Markus Q.~Huber}
\affiliation{Institute of Physics, University of Graz, NAWI Graz, Universit\"atsplatz 5, 8010 Graz, Austria}
\email{markus.huber@uni-graz.at}
\date{\today}

\begin{abstract}
We compute the gluon and ghost propagators of Yang-Mills theory in linear covariant gauges from the coupled system of Dyson-Schwinger equations. For small values of the gauge fixing parameter $\xi\sim 0.1$, the deviations to the Landau gauge already become clearly visible. For the ghost dressing function, this is reflected in a logarithmic infrared suppression. Also, the gluon dressing function changes -- but only quantitatively -- and the gluon propagator remains finite at zero momentum. From the results, a running coupling is extracted.
\end{abstract}



\maketitle

\section{Introduction}

Quantum chromodynamics is a theory rich with physical phenomena, e.g., asymptotic freedom, confinement, dynamical mass creation, to name only a few well-known examples. The investigation of its low energy regime faces the challenge of being nonperturbative and requires corresponding methods like lattice simulations or functional methods.

In this paper we will investigate Yang-Mills theory -- i.e., we neglect all quark effects -- with Dyson-Schwinger equations (DSEs). The basic quantities in this approach are correlation functions of fields. The lowest ones, the propagators, are particularly important, as they encode most of the nonperturbative dynamics.
As correlation functions are gauge dependent quantities, the choice of a gauge is required. This allows, to some extent, to avoid or alleviate some difficulties. For example, different gauges feature a different number of fields and/or interactions. Thus gauges with a minimal number of both have a technical advantage over others. Indeed, the gauge which was most prominently used during the last decades fulfills this criterion: The Landau gauge has two fields (gluon and ghost) and three primitively divergent vertices. 
Calculations in this gauge have advanced as far as the self-consistent calculation of its primitively divergent vertices~\cite{Huber:2012kd,Blum:2014gna,Eichmann:2014xya,Cyrol:2014kca}. 
Of course there are also other reasons for the attractiveness of the Landau gauge. One is the fact that the transverse correlation functions form a closed system and decouple from the longitudinal ones \cite{Fischer:2008uz}. Historically the simple form of the  ghost-gluon vertex played an important role as it provided the entry point to the coupled system of gluon and ghost propagators \cite{vonSmekal:1997vx}. Since then, much work on the propagators has been done; see, for example, \cite{vonSmekal:1997vx,Atkinson:1997tu,Alkofer:2000wg,Zwanziger:2001kw,Lerche:2002ep,Fischer:2002hn,Pawlowski:2003hq,Fischer:2006ub,Alkofer:2008jy,Aguilar:2008xm,Fischer:2008uz,Huber:2009tx,Strauss:2012dg,Huber:2012kd,Blum:2014gna}.

Other gauges besides the Landau gauge have also been used in the past. Among them, the Coulomb gauge is the gauge where the study of the elementary Green functions has also progressed rather far \cite{Feuchter:2004mk,Campagnari:2010wc,Campagnari:2011bk,Huber:2014isa}. Another example is the maximal Abelian gauge, the correlation functions of which have also been studied to some extent, e.g., \cite{Capri:2008ak,Huber:2009wh}. And, last but not least, there are the linear covariant gauges, the end point of which is the Landau gauge. Within the functional framework they have been investigated only occasionally, e.g., \cite{Alkofer:2003jr,Aguilar:2007nf}.
Results from different gauges offer the possibility of testing the gauge (in)dependence of observables. Naturally, observables are gauge independent, but truncating the underlying functional equations can spoil this property. In this respect linear covariant gauges play a special role, as they allow us to change the gauge continuously and have the Landau gauge as their end point, which is well studied.

At first sight, the extension to a nonvanishing value of the gauge fixing parameter of the linear covariant gauges, $\xi$, may seem rather straightforward once the necessary techniques are mastered. However, the little knowledge we have about this family of gauges for providing a guide in the construction of a truncation for functional equations is mostly based on an extrapolation from the Landau gauge. In addition, the longitudinal parts of the correlation functions have to be considered.

In this context it is interesting to note that with lattice calculations the extension to nonzero $\xi$ is also a nontrivial issue. However, the main reason for this is the fact that conventional gauge fixing techniques are based on the extremization of a functional; see \eref{eq:RA} below. Such functionals exist for the Landau gauge, the maximal Abelian gauge, and the Coulomb gauge, but for linear covariant gauges one can show that such a functional cannot be constructed in the conventional way and that alternative methods are required \cite{Giusti:1996kf,Giusti:2000yc,Cucchieri:2008zx,Cucchieri:2009kk}; see Ref.~\cite{Cucchieri:2011pp} for an overview on this topic.

Here we present an exploratory investigation of the propagators in linear covariant gauges using Dyson-Schwinger equations. The truncation we use is motivated by its successful application in the Landau gauge. Lacking information on the transverse parts of the vertices we construct models that smoothly connect to the Landau gauge but contain only the minimally required ingredients like the correct ultraviolet (UV) behavior. For the longitudinal parts we rely on information provided by Ward identities. The resulting system of equations is then solved self-consistently.

The article is organized as follows.
Yang-Mills theory in linear covariant gauges is discussed in Sec.~\ref{sec:YM-LinCov}. The Dyson-Schwinger equations for the propagators are introduced in Sec.~\ref{sec:DSEs}, where the IR behavior of the ghost propagator is also investigated. The numerical results are presented in Sec.~\ref{sec:results} and in Sec.~\ref{sec:summary} we summarize. In the Appendix some details of the Dyson-Schwinger equations are discussed.

\section{Linear covariant gauges}
\label{sec:YM-LinCov}

The action of Yang-Mills theory, $S_{\text{YM}}=\frac{1}{4}\int d^4x F_{\mu \nu }^aF^a_{\mu \nu}$, 
is invariant under gauge transformations of the gauge field $A_\mu^a$,
\begin{align}
A_\mu^a \rightarrow A_\mu^a-D_\mu^{ab} \theta^b, \quad D_\mu^{ab}=\delta^{ab}\partial_\mu+g\,f^{abc} A_\mu^c.
\end{align}

The Landau gauge is defined as minimizing the norm of the gauge fields
\begin{align}\label{eq:RA}
R[A]=\int d^4x\, A_\mu^a A_\mu^a
\end{align}
with respect to  gauge transformations. Any gauge field configuration obeying $\partial_\mu A_\mu^a=0$ fulfills this criterion. However, nonperturbatively this gauge fixing is not unique and the issue of Gribov copies arises \cite{Vandersickel:2012tz}. The restriction to the hyperplane $\partial_\mu A_\mu^a=0$ introduces auxiliary fields called ghosts via localization of the Jacobian determinant:
\begin{align}
S_{\text{gf}}&=\int d^4x\, \left(\frac1{2\xi}(\partial_\mu A_\mu^a)^2-\bar{c}^a M^{ab} c^b) \right).
\end{align}
In the limit $\xi=0$ the condition $\partial_\mu A_\mu^a=0$ is strictly enforced. This is the case of the Landau gauge. $\xi>0$ corresponds to a Gaussian distribution along the gauge orbit centered at $\partial_\mu A_\mu^a=0$ with a width of $\xi$. Gribov copies manifest then in the existence of several such Gaussians.

The dressed gluon propagator is parametrized as
\begin{align}
 D_{\mu\nu}^{ab}(p)&=D_{\mu\nu}^{ab,T}(p)+\xi \de^{ab}\frac{p_\mu p_\nu}{p^4}
\end{align}
with
\begin{align}
D_{\mu\nu}^{ab,T}(p)=\frac{\delta^{ab}}{p^2}Z(p^2)P_{\mu\nu}(p)
\end{align}
being the transverse part. $P_{\mu\nu}(p)=g_{\mu\nu}-p_\mu p_\nu/p^2$ is the transverse projector. As is known from the Slavnov-Taylor identity for the gluon propagator, the longitudinal part stays bare. In perturbative calculations, where $Z(p^2)=1$ is used, the Feynman gauge with $\xi=1$ simplifies the calculations, but in a nonperturbative setting the dressing of the transverse part negates that and the Feynman gauge is not simpler. Another peculiar value for $\xi$ is $3$, for which the ghost self-energy becomes finite; see Tab.~\ref{tab:anom_dims}. This is the Yennie gauge. The unitary gauge is $\xi\rightarrow \infty$.

\begin{table}[tb]
 \begin{tabular}{|l|c|}
 \hline
 Green function & Anomalous dimension \\
 \hline \hline
  Ghost propagator & $\delta=-\frac{9-3\xi}{44} $\\
  \hline
  Gluon propagator & $\gamma=-\frac{13-3\xi}{22}$\\
  \hline
  Ghost-gluon vertex & $\gamma^{ghg}=-\frac{3\xi}{22}$\\
  \hline
  Three-gluon vertex & $\gamma^{\text{3g}}=\frac{17-9\xi}{44}$\\
  \hline
 \end{tabular}
\caption{\label{tab:anom_dims}The anomalous dimensions of the propagators and vertices.}
\end{table}

The inverse of the propagator, the two-point function, is written as
\begin{align}
 \Gamma_{\mu\nu}^{ab}(p)&=\left(\Gamma_{\mu\nu}^{ab,T}(p)+\xi^{-1} \de^{ab}p_\mu p_\nu\right)
\end{align}
with
\begin{align}
\Gamma_{\mu\nu}^{ab,T}(p)&=\delta^{ab}Z^{-1}(p^2)P_{\mu\nu}(p) p^2.
\end{align}

The ghost propagator $D^{ab}(p)$ is parametrized by
\begin{align}
 D^{ab}(p)&=-\delta^{ab}G(p^2)\frac1{p^2}.
\end{align}

The two dressing functions of the ghost-gluon vertex are taken as the parts transverse and longitudinal with respect to the gluon leg:
\begin{align}\label{eq:ghg-tr-long}
&\Gamma_\mu^{abc}(k;p,q)=\nnnl
&i\,g f^{abc}\left(D^{\text{gg}}_T(k^2;p^2,q^2)P_{\mu\nu}(k)p_\nu+D^{\text{gg}}_L(k^2;p^2,q^2)k_\mu\right).
\end{align}
The momenta $k$, $p$, $q$ are those of the gluon, the antighost and the ghost. The bare vertex $\Gamma_\mu^{abc,(0)}(k;p,q)$ is obtained by setting $D^{\text{gg}}_T(k^2;p^2,q^2)=1$ and $D^{\text{gg}}_L(k^2;p^2,q^2)=p\cdot k/k^2$.
In the Landau gauge the ghost-gluon vertex is often used without radiative corrections. In particular, it is known that it is UV finite \cite{Taylor:1971ff}. This is no longer true for $\xi>0$ and we implement the running of the vertex via the following model:
\begin{align}
 D^{\text{gg}}_T(k^2;p^2,q^2)=G(\bar{p}^2)^{\alpha_1^{\text{gg}}} Z(\bar{p}^2)^{\beta_1^{\text{gg}}}.
\end{align}
$\bar{p}^2$ is $(p^2+q^2+k^2)/2$. The exponents $\alpha_1^{\text{gg}}$ and $\beta_1^{\text{gg}}$ can be found in Tab.~\ref{tab:exps_RGI} in the Appendix where their calculation is also discussed since they are the same ones used for the renormalization group improvement. Using Tab.~\ref{tab:anom_dims}, it can be easily checked that this expression has the correct UV behavior.

The full three-gluon vertex has 14 tensors, of which four form the transverse subspace. Motivated by results in the Landau gauge, where it was shown by explicit calculations that all dressing functions of the transverse tensors are negligible except for the contribution from the tree-level tensor \cite{Eichmann:2014xya}, we restrict ourselves to this tensor given by
\begin{align}
&\Gamma_{\mu\nu\rho}^{abc,(0)}(p,q,r)=\,i\,g\,f^{abc}\left( (r-q)_\mu g_{\nu\rho} + \text{perm.} \right).
\end{align}
The transverse part of the full three-gluon vertex is then written as the transversely projected bare tensor with a dressing $D^{\text{3g}}(k^2,p^2,q^2)$ given by
\begin{align}
 D^{\text{3g}}(k^2,p^2,q^2)=G(\bar{p}^2)^{\alpha_1^{\text{3g}}} Z(\bar{p}^2)^{\beta_1^{\text{3g}}}.
\end{align}

For the longitudinal parts of the vertices we use Ward identities. They are obtained from the invariance of the path integral under gauge transformations, see Refs.~\cite{Litim:1998qi,Freire:2000bq,Pawlowski:2005xe,Gies:2006wv} for details:
\begin{align}
\frac1{Z}\int D\Phi \, \mathcal{G}^a e^{-S_{\text{YM}}-S_{\text{gf}}-S_{\text{gh}}-S_{\text{sources}}}=0,
\end{align}
where
\begin{align}
\mathcal{G}^a=D_\mu^{ab}(x)\frac{\de}{\de A_\mu^b(x)}+g\,f^{abc}\left(c^c \frac{\de}{\de c^b}+\bar{c}^c \frac{\de}{\bar{c}^b} \right)
\end{align}
is the Ward operator. From this identity we calculate the Ward identities for the three-point functions by applying further derivatives. As an approximation we keep only the terms of lowest order in $g$. The resulting expressions are then used for the longitudinally projected vertices in the DSEs.
For the ghost-gluon vertex we obtain
\begin{align}\label{eq:ghg_WI}
 D^{\text{gg}}_L(k^2;p^2,q^2)=\frac{q^2\,G^{-1}(q^2)-p^2\,G^{-1}(p^2)-k^2-p\cdot k}{k^2}
\end{align}
and for the three-gluon vertex
\begin{align}
i\,p_\mu\,\Gamma_{\mu\nu\rho}^{abc}(p,q,r)=g\,f^{abc}\left(\Gamma_{\nu\rho}^{T}(r)-\Gamma_{\nu\rho}^{T}(q)\right).
\end{align}

\section{The propagator Dyson-Schwinger equations}
\label{sec:DSEs}

\begin{figure}[tb]
\includegraphics[width=0.35\textwidth]{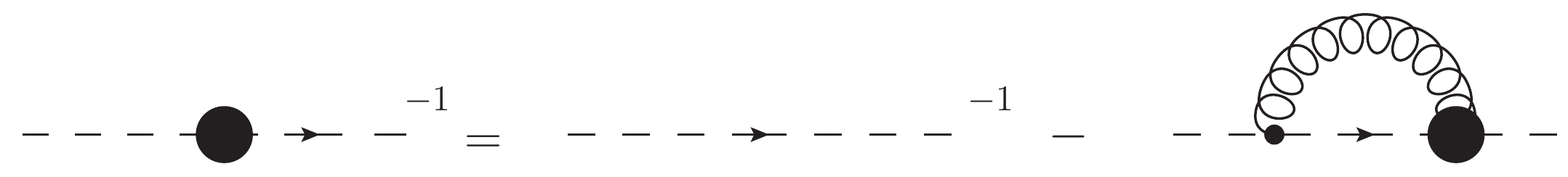}\phantom{XXXXXXXX}\\
\vskip5mm
\includegraphics[width=0.45\textwidth]{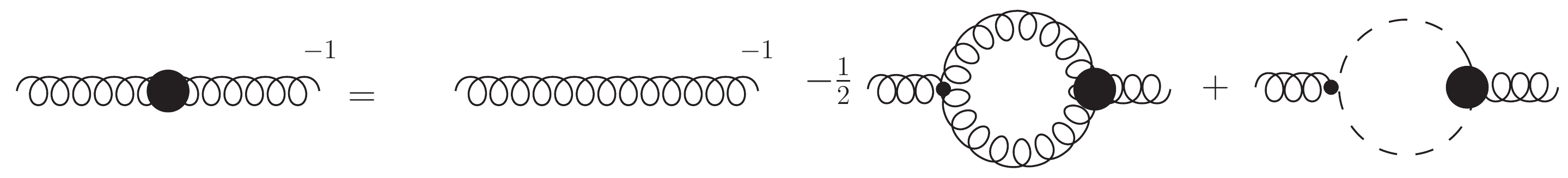}
\caption{\label{fig:DSEs}Truncated two-point Dyson-Schwinger equations. All internal propagators are dressed. Thick blobs denote dressed vertices. Wiggly lines are gluons, dashed ones ghosts. Plots were created with \textit{Jaxodraw} \cite{Binosi:2003yf}.}
\end{figure}

Based on the invariance of the path integral under translations of the fields, the equations of motion for all correlation functions can be derived; see, e.g., \cite{Alkofer:2000wg,Alkofer:2008nt} for details. On a formal level the DSEs in linear covariant gauges look the same as in the Landau gauge, viz.,
\begin{widetext}
\begin{align}
\label{eq:gh-DSE}
 \left(D^{ab}(p)\right)^{-1}&=-\widetilde{Z}_3 p^2+g^2\,\tilde{Z}_1\int_q D^{cd}(p+q) D^{ef}_{\mu\nu}(q)\Gamma_\nu^{fdb}(-q;p+q,-p)\Gamma_\mu^{eac,(0)}(q;p,-q-p)\\
\label{eq:gl-DSE}
 \left(D^{ab}_{\mu\nu}(p)\right)^{-1}&= \delta^{ab}P_{\mu\nu}(p)p^2Z_3+\xi^{-1} \de^{ab}p_\mu p_\nu+g^2\,\tilde{Z}_1\int_q D^{cd}(q)D^{ef}(p+q) \Gamma^{bfd}_\nu(-p;p+q,-q)\Gamma^{ace,(0)}_\mu(p;q,-p-q)\nnnl
 &-\frac{g^2}{2}\,Z_1\int_q D^{cd}_{\rho\sigma}(q)D^{ef}_{\alpha\beta}(p+q) \Gamma^{bfd}_{\nu\beta\sigma}(-p,p+q,-q)\Gamma^{ace,(0)}_{\mu\rho\alpha}(p;q,-p-q)+\ldots,
\end{align}
\end{widetext}
see also \fref{fig:DSEs}.
$\int_q$ stands for $\int d^4q/(2\pi)^4$ and $\widetilde{Z}_1$ and $Z_1$ are the renormalization constants of the ghost-gluon and three-gluon vertices, respectively, whereas $\tilde{Z}_3$ is the renormalization constant of the ghost propagator and $Z_3$ that for the gluon propagator. The dots represent two-loop terms and the tadpole diagram. The latter is typically neglected as it does not contribute perturbatively and its nonperturbative impact is small at best~\cite{Huber:2014tva}. The two-loop terms, on the other hand, are important in the midmomentum regime, at least in the Landau gauge \cite{Blum:2014gna,Meyers:2014iwa}. However, for the moment we leave these contributions aside. Before we can aim at quantitative results we would need more detailed information about the vertices in general, which is not available.

We bring the DSEs now into a form amenable to numerical computation.
We start with the ghost propagator DSE (\ref{eq:gh-DSE}).
Plugging in explicit expressions for the propagators and vertices, it becomes
\begin{widetext}
\begin{align}
 G(x)^{-1}&=\widetilde{Z}_3+\frac{g^2 \, N_c\,\widetilde{Z}_1}{4}\int_q \frac{G(z)}{x\, y^2\, z}\left((x^2+(y-z)^2-2x(y+z))Z(y)D_T^{\text{gg}}(y,z,x)-2\xi\,y(x+y-z)D_L^{\text{gg}}(y,z,x) \right).
\end{align}
Here and in the following, we use $x=p^2$, $y=q^2$ and $z=(p+q)^2$. Switching to hyperspherical coordinates, we can integrate out two angles and the integrand reduces to two terms:
\begin{align}
 G(x)^{-1}&=\widetilde{Z}_3+\frac{g^2\,N_c}{4\pi^3} \int dq \,d\varphi \frac{q\sin(\varphi)^2 }{z\,Z_c(z)}\left(-\frac{\sin(\varphi)^2  D^{A\bar c c}_T(y,z,x)}{Z_A(y)}+\xi\frac{q \cos \varphi}{p}D^{A\bar c c}_L(y,z,x) \right).
\end{align}
The final expression is obtained by plugging in the Ward identity for the ghost-gluon vertex from \eref{eq:ghg_WI}:
\begin{align}\label{eq:gh-DSE-final}
 G^{-1}(x)=\widetilde{Z}_3+\frac{g^2\,N_c\,\widetilde{Z}_1}{8\pi^3} \int dy\,d\varphi \frac{\sin(\varphi)^2
}{z}G(z)
\Bigg(-\sin(\varphi)^2 D^{A\bar c c}_T(y,z,x)Z(y)
+\xi\, \frac{p\, \cos \varphi}{q \,G(x)}+\xi\,\cos(\varphi)^2 \Bigg),
\end{align}
where one term vanished due to the angle integration.
\end{widetext}

From this expression the infrared (IR) behavior of the ghost propagator can be inferred. As we will argue, the term proportional to $\xi$ is IR leading, which entails a different IR behavior from the Landau gauge. In the Landau gauge, only the first term appears, which goes to zero for $p\rightarrow 0$ since the product $G(z)Z(y)D^{A\bar c c}_T(y,z,x)$ vanishes faster than the rest of the integrand. If that were different for $\xi>0$, the qualitative IR behavior of the propagators or the ghost-gluon vertex would need to change drastically. We will assume that this is not the case and will support this by an explicit calculation. For low external momentum $p$, the third part becomes
\begin{align}
\xi\frac{g^2\,N_c\,\widetilde{Z}_1}{8\pi^3} \int dy\,d\varphi \frac{\sin(\varphi)^2
}{y}\cos(\varphi)^2 G(y).
\end{align}
If $G(y)$ were IR divergent or constant this would lead to an IR divergence in the integral. In turn this would mean that $G(x)$ on the left-hand side would vanish in contradiction to the original assumption. Thus the ghost dressing function must be IR suppressed to make the integrand convergent. The numerical results will indeed show this behavior.

\begin{figure*}[tb]
 \includegraphics[width=0.48\textwidth]{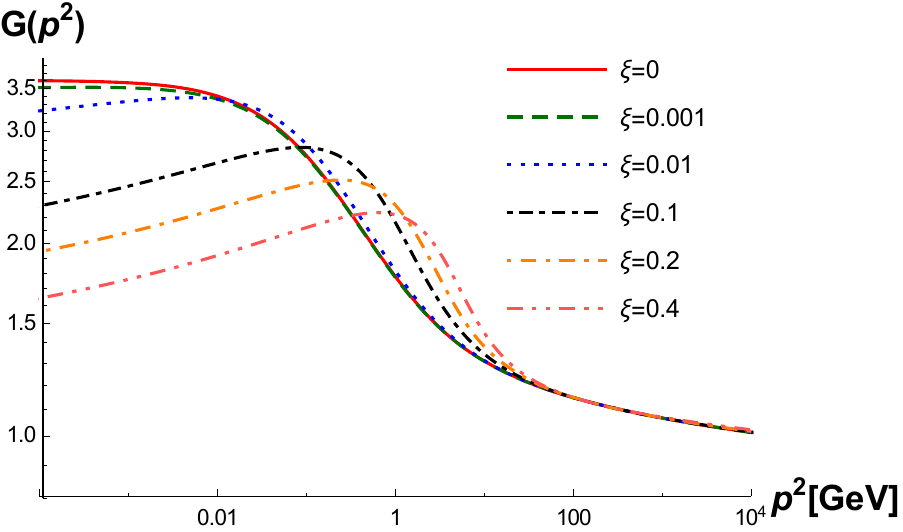}
 \hfill
 \includegraphics[width=0.48\textwidth]{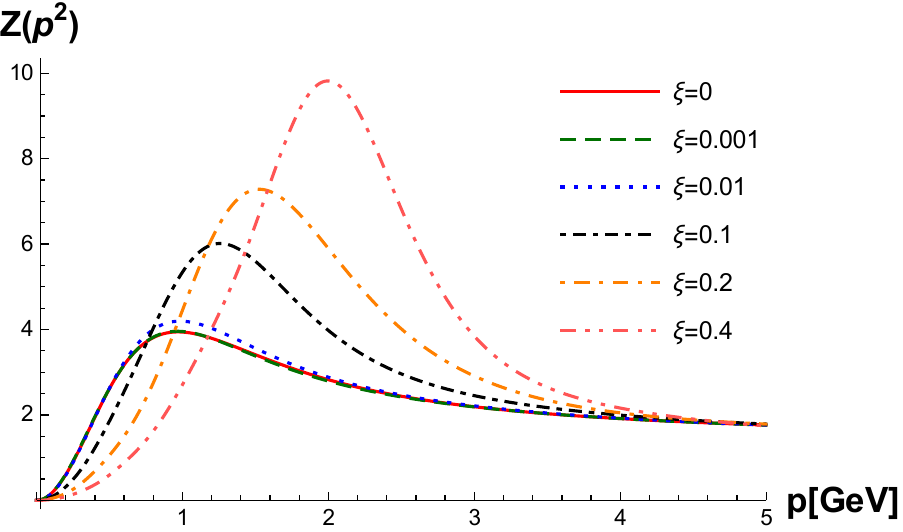}
 \caption{\label{fig:res_props}Ghost (left) and gluon (right) dressing functions for various values of $\xi$.}
\end{figure*}

We now turn to the gluon propagator DSE.
To transform it into a scalar equation we project it with the transverse projector. Note that then only the transverse part of the ghost-gluon vertex appears. Splitting the gluon loop by orders in $\xi$, the DSE reads
\begin{align}\label{eq:gl-DSE-final}
 Z&^{-1}(p^2)=\nnnl
 &Z_3+g^2\,N_c\,\tilde{Z}_1\int_q\,  G(y)G(z) K_{Z}^{gh}(x;y,z)D_T^{\text{gg}}(x;y,z)\nnnl
 &+g^2\,N_c\,Z_1\,\int_q \Big( Z(y)Z(z) K_{Z}^{gl}(x;y,z)D^{\text{3g}}(x,y,z) \nnnl&
 \quad+ \xi K_Z^{gl,\xi} +
 \xi^2 K_Z^{gl,\xi^2} Z(x)^{-1}\Big).
\end{align}
The kernels are given in \eref{eq:kernels} in the Appendix.
Note that in $K_Z^{gl,\xi}$ the gluon dressing function appears in various combinations.

Before the system of equations can be solved some further modifications are required. They are related to the UV behavior of the equations, which should consistently produce the one-loop resummed perturbation theory. However, the truncation spoils this property. Since the employed Newton method is sensitive to a consistent UV behavior, we restore this property by adding renormalization group improvement terms \cite{Huber:2012kd}. The procedure on how to do this is described in the Appendix. The issue of spurious UV divergences which are subtracted perturbatively is also discussed there~\cite{Huber:2014tva}. The final equations including these modifications can be found in the Appendix in Eqs.~(\ref{eq:gh-DSE-final-wRGI}) and (\ref{eq:gl-DSE-final-wRGI}).

An interesting question with regard to the IR behavior of Green functions is if a scaling solution exists also for nonzero $\xi$. Such a solution is characterized by power laws for the dressing functions \cite{vonSmekal:1997vx,Fischer:2009tn}. A DSE analysis using a bare vertex approximation found that this is not possible if one requires that the longitudinal gluon propagator stays bare \cite{Alkofer:2003jr}. This also holds if dressed vertices are considered \cite{Huber:2010ne}. The only way out would be nontrivial cancelations in the loop diagrams, which, however, we do not find here. However, given the analytic result that the ghost dressing function is IR suppressed, it is hard to see how a scaling solution could be realized.

\section{Results}
\label{sec:results}

\begin{figure*}[tb]
 \includegraphics[width=0.48\textwidth]{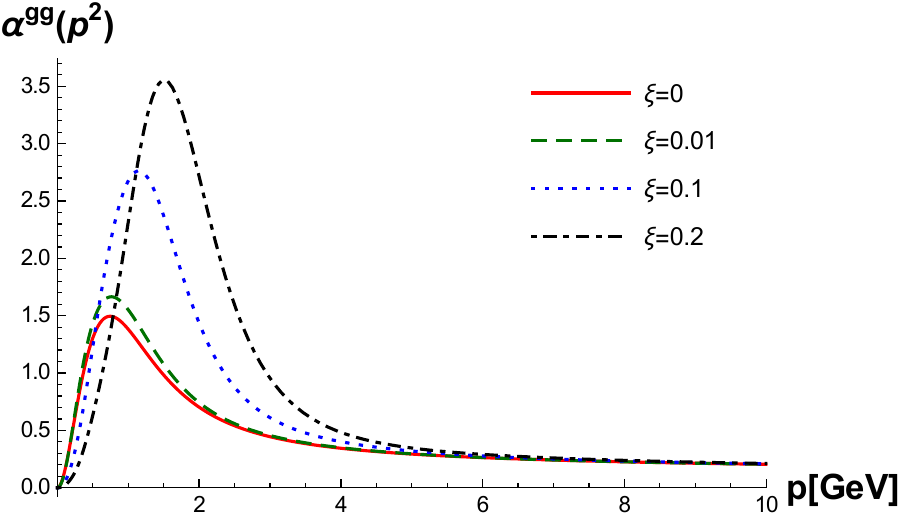}
 \hfill
 \includegraphics[width=0.48\textwidth]{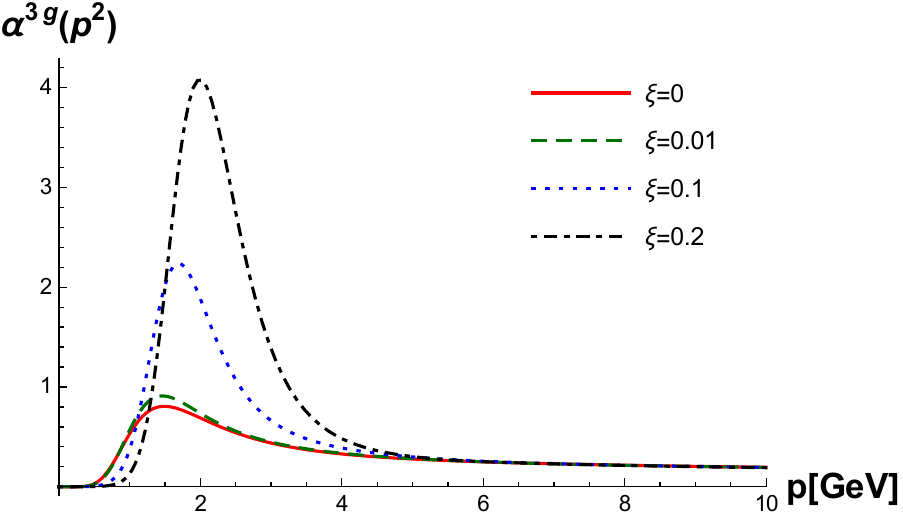}
 \caption{\label{fig:res_coups}Running couplings from the ghost-gluon (left panel) and the three-gluon (right panel) vertices.}
\end{figure*}

Using the framework provided by \textit{DoFun} \cite{Huber:2011qr,Alkofer:2008nt} and \textit{CrasyDSE} \cite{Huber:2011xc} the system of the DSEs~(\ref{eq:gh-DSE-final-wRGI}) and (\ref{eq:gl-DSE-final-wRGI}) is solved using a standard Newton method, see, e.g., \cite{Atkinson:1997tu,Maas:2005xh,Huber:2011qr}. Solutions for various values of $\xi$ including $0$ are determined. It turned out that for a nonzero gauge fixing parameter, a higher precision is required and that with the currently employed setup the value of $\xi$ is limited to $\xi\lesssim0.4$. For larger values, numerical artifacts become too large.

Our results are shown in \fref{fig:res_props}. Even though Yang-Mills theory does not have a physical scale, one can use the string tension or quark potential methods to set one artificially. To inherit this scale, we use the distinct bump in the Landau gauge gluon dressing function as a marker. Using truncation schemes including the vertices one can obtain with this method very good agreement with the running coupling in the universal perturbative regime \cite{Huber:2012kd}. Here we used lattice results from Ref.~\cite{Sternbeck:2006rd} and calculated the scale setting factor from the Landau gauge result. It was then applied for all values of $\xi$.

First differences to the Landau gauge are seen in the ghost dressing function which starts to bend down in the IR. This effect, which can be seen already for very small values of the gauge fixing parameter, $\xi\sim 0.001$, where the gluon dressing function is still unaffected, is in agreement with the qualitative IR analysis in Sec.~\ref{sec:DSEs}. The IR suppression is logarithmic and not as strong as $p^2$ which would make the ghost propagator IR finite. This behavior was found in an earlier DSE analysis \cite{Aguilar:2007nf}. In another study using a variational method the ghost dressing function was found to be constant at low momenta \cite{Siringo:2014lva}. Around $\xi=0.1$ the gluon dressing also starts to change and the bump in the midmomentum regime becomes bigger. At this value of the gauge fixing parameter, the ghost dressing function deviates in the IR already severely from the Landau gauge. Raising $\xi$ further, the ghost dressing function becomes even more IR suppressed. The effect in the gluon dressing is most drastic in the midmomentum regime, where the bump gets enlarged. At small momenta the gluon dressing function always vanishes like $p^2$, so that the propagator is IR finite. This is in agreement with the result of Ref.~\cite{Siringo:2014lva}.
The UV behavior for both propagators changes as expected. 

While for small values of $\xi$ the proximity to the well-known Landau gauge most likely makes the results trustworthy, it is unclear how well the employed approximations work for larger values of $\xi$. In particular, the shift of the bump in the gluon dressing function to higher momenta and the large increase in height indicates that the employed truncation may still lack some important features. An obvious possibility for improving the truncation is to use more realistic vertex dressings. The employed models are rather simple and do not capture any nontrivial structures in the nonperturbative regime.

We also calculate the running couplings as extracted from the ghost-gluon and three-gluon vertices \cite{vonSmekal:1997vx,Alkofer:2004it}:
\begin{align}
 \alpha^{\text{gg}}(p^2) & = \alpha(\mu^2) \, \left[ D^{\text{gg}}(p^2) \right]^2 G^2(p^2) Z(p^2),\\
 \alpha^{\text{3g}}(p^2) & = \alpha(\mu^2) \, \left[ D^{\text{3g}}(p^2) \right]^2 Z^3(p^2).
\end{align}
Note that the ghost-gluon vertex enters explicitly, in contrast to the Landau gauge, where it is UV finite. The results are shown in \fref{fig:res_coups}. In the perturbative regime they all agree as expected from universality. The large bump in the gluon dressing function leads to a rise at intermediate momenta for a larger $\xi$. For increasing values of $\xi$, the bump in the coupling also moves to higher values. Unfortunately, our results do not show any sign of a slowing down of that movement, so that for large enough values of $\xi$ this will be in conflict with perturbative universality. This hints again at a shortcoming of the vertex models for larger values of $\xi$.

Finally, we note that a desired property of our solution is that it fulfills a confinement criterion based on the Polyakov loop potential \cite{Braun:2007bx,Fister:2013bh}. The important property is that the gluon propagator is IR suppressed relative to the ghost propagator, which then leads to a confining Polyakov loop potential.

\section{Summary and conclusions}
\label{sec:summary}

In this work we presented results for the propagators of Yang-Mills theory in linear covariant gauges from Dyson-Schwinger equations. The truncation was kept quite simple, with the only dynamic quantities being the propagators. For the vertices, models were used for the transverse parts that respect the correct UV behavior and the longitudinal parts were taken from their Ward identities in leading order in $g$. From the results for the propagators a running coupling was extracted.

Analytically, we could show from the ghost propagator DSE that the ghost dressing function must be IR suppressed. The numerical results confirmed that behavior. Thus the ghost propagator has a different IR behavior from the Landau gauge. The gluon propagator, on the other hand, qualitatively has the same IR behavior; viz., it becomes constant. With an increasing value of the gauge fixing parameter $\xi$, the bump in the gluon dressing function shifts towards larger momenta and becomes higher. However, this may be a truncation artifact. To investigate this issue further dedicated investigations of the vertices -- or at least the use of improved models -- would be required.

\textit{Note added:}
Recently Ref.~\cite{Aguilar:2015nqa} appeared, where the propagators of linear covariant gauges were investigated. The numerical results obtained for the ghost propagator are in qualitative agreement with ours.

\begin{acknowledgments}
During this project I profited greatly from discussions with Reinhard Alkofer and Jan M. Pawlowski about linear covariant gauges. In particular the idea of how to use the Ward identities in this context can be attributed to Jan M. Pawlowski and I am grateful to him and Sergei Nedelko for showing me related unpublished material.
Funding by the FWF (Austrian science fund) under Contract P 27380-N27 is gratefully acknowledged.
\end{acknowledgments}

\vskip5mm

\appendix

\setcounter{secnumdepth}{0}

\section{Appendix: Details on the Dyson-Schwinger equations of the propagators}

The kernels for the gluon propagator DSE in \eref{eq:gl-DSE-final} are given by 
\begin{widetext}
\begin{subequations}\label{eq:kernels}
\begin{align}
K_{Z}^{gh}(x;y,z)&=-\frac{  \left(x^2-2 x (y+z)+(y-z)^2\right) }{12 x^2 y z},\\
K_{Z}^{gl}(x;y,z)&=\frac{ \left(x^2-2 x (y+z)+(y-z)^2\right) \left(x^2+10 x (y+z)+y^2+10 y z+z^2\right)}{24 x^2 y^2 z^2},\\
K_Z^{gl,\xi}(x;y,z)&=\frac{  \left(x^3 (y+z)+x^2 \left(9 y^2-4 y z+9 z^2\right)+x \left(-9 y^3+y^2 z+y z^2-9 z^3\right)-(y-z)^2 \left(y^2+z^2\right)\right)}{24 x^2 y^2 z^2}\nnnl
&-\frac{Z(z) (x-z) \left(x^2-2 x (y-5 z)+(y-z)^2\right)}{24 x y^2 z^2Z(x)}
 -\frac{  Z(y) (x-y) \left(x^2+2 x (5 y-z)+(y-z)^2\right)}{24 x y^2 z^2 Z(x)},\\
 K_Z^{gl,\xi^2}(x;y,z)&=\frac{ \left(x^2-2 x (y+z)+(y-z)^2\right)}{24 y^2 z^2}.
\end{align}
\end{subequations}
\end{widetext}

To solve the system of equations, a standard Newton method is employed. A stable iteration is only achieved if the anomalous dimensions are reproduced self-consistently. However, one-loop resummed perturbation theory requires a resummation of the diagrams not included in the truncation. In the Landau gauge, one possible remedy is to modify the vertex models such that the correct running is produced, or, in other words, to use momentum dependent renormalization constants \cite{vonSmekal:1997vx,Huber:2012kd,Huber:2014tva}. Because of the presence of the mixed terms in the kernel $K_Z^{gl,\xi}$, this procedure needs to be modified: Every term gets multiplied by a so-called renormalization group improvement factor to obtain the correct one-loop running \cite{Huber:2012kd}. This factor is generically
\begin{align}
 F(\alpha,\beta;\bar{p}^2)=G(\bar{p}^2)^\alpha Z(\bar{p}^2)^\beta,
\end{align}
where $\bar{p}^2$ is $(x+y+z)/2$. This choice ensures that for large loop momenta $q$ the argument becomes $y$ while being symmetric in all momenta. The exponents $\alpha$ and $\beta$ are determined such that the logarithmic running in the UV is correct. In addition we would require, as in the Landau gauge, that $F(\alpha,\beta; \bar{p}^2)$ would become constant in the IR if the ghost propagator were IR constant. However, since the ghost propagator is not IR constant the latter condition should be improved in future calculations, particularly as it can have a quantitative influence on the results \cite{Eichmann:2014xya,Huber:2014tva}. However, we adopt it here due to its simplicity. The final DSEs then read
\begin{widetext}
\begin{align} 
\label{eq:gh-DSE-final-wRGI}
G^{-1}(x)&=\widetilde{Z}_3+\frac{g^2\,N_c}{8\pi^3} \int dy\,d\varphi \frac{\sin(\varphi)^2
}{z}\Bigg(-\sin(\varphi)^2 D^{A\bar c c}_T(y,z,x)Z(y)F(\alpha_1^{\text{gg}},\beta_1^{\text{gg}};\bar{p}^2)\nnnl
&+\xi F(\alpha_2^{\text{gg}},\beta_2^{\text{gg}};\bar{p}^2)\left( \frac{p\, \cos \varphi}{q \,G(x)}+\cos(\varphi)^2 \right)\Bigg)G(z),\\
\label{eq:gl-DSE-final-wRGI}
 Z^{-1}(p^2)&=Z_3+g^2\,N_c\,\int_q\,  G(y)G(z) K_{Z}^{gh}(x;y,z)D_T^{\text{gg}}(x;y,z)F(\alpha_1^{\text{gg}},\beta_1^{\text{gg}},\bar{p}^2)\nnnl
 &+g^2\,N_c\,\int_q \left( Z(y)Z(z) K_{Z}^{gl}(x;y,z)D^{\text{3g}}(x,y,z)F(\alpha_1^{\text{3g}},\beta_1^{\text{3g}};\bar{p}^2) +
 \xi \widetilde{K}_Z^{gl,\xi} +
 \xi^2 K_Z^{gl,\xi^2} Z(x)^{-1}\right)
\end{align}
with the new kernel given by
\begin{align} 
\widetilde{K}_Z^{gl,\xi}(x;y,z)&=\frac{  \left(x^3 (y+z)+x^2 \left(9 y^2-4 y z+9 z^2\right)+x \left(-9 y^3+y^2 z+y z^2-9 z^3\right)-(y-z)^2 \left(y^2+z^2\right)\right)}{24 x^2 y^2 z^2}F(\alpha_3^{\text{3g}}, \beta_3^{\text{3g}}; \bar{p}^2)\nnnl
&-\frac{Z(z) (x-z) \left(x^2-2 x (y-5 z)+(y-z)^2\right)}{24 x y^2 z^2Z(x)}Z(x)F(\alpha_2^{\text{3g}}, \beta_2^{\text{3g}}; \bar{p}^2)\nnnl
 &-\frac{  Z(y) (x-y) \left(x^2+2 x (5 y-z)+(y-z)^2\right)}{24 x y^2 z^2 Z(x)}Z(x)F(\alpha_2^{\text{3g}}, \beta_2^{\text{3g}}; \bar{p}^2).
\end{align}
\end{widetext}
The renormalization constants have all been dropped in favor of the renormalization group improvement terms.
Note that the renormalization group improvement terms for two terms also include an additional factor $Z(x)$.
The values for the exponents are given in Tab.~\ref{tab:exps_RGI}.

A particular problem of the gluon propagator DSE is the appearance of spurious divergences. There are several ways to get rid of them. We adopt the one from Ref.~\cite{Huber:2014tva} here. There it was shown that their origin is purely perturbative. Consequently we can calculate them analytically and use these expressions to subtract them. The generic structure of the subtraction term is $C_\mathrm{sub}/p^2$, with
\begin{align}\label{eq:C_sub}
C_\mathrm{sub}&=
\LQ \,b\,\omega^{-1-\gamma}\sum_{n=0}^{\infty}\frac{\left(\ln\left(\LL/\LQ\right)\right)^{-\gamma+n}}{n!(-\gamma+n)}
\end{align}
where $\Lambda$ is the UV cutoff and $\Lambda_\mathrm{QCD}$ is defined as the position of the one-loop Landau pole, viz., $\Lambda^2_\mathrm{QCD}=s\,e^{-1/\omega}$. $\omega$ is given by $11\,N_c\,\alpha(s)/12/\pi$, where $\alpha(s)$ is the running coupling at a perturbative scale $s$. The coefficient $b$ is determined from the high momentum behavior in the gluon propagator DSE:
\begin{align}
 b&=\frac{g^2\,N_c}{64\pi^2}\Big( G(s)^{2+2\alpha_1^{\text{gg}}} Z(s)^{2\beta_1^{\text{gg}}}\nnnl
 &- 6 G(s)^{2\alpha_1^{\text{3g}}} Z(s)^{2+2\beta_1^{\text{3g}}}
 - 3\xi G(s)^{\alpha_3^{\text{3g}}} Z(s)^{\beta_3^{\text{3g}}}
 \Big).
\end{align}
Again the part proportional to $\xi^2$ does not contribute.
To deal with the logarithmic divergences a MOM scheme is used where the dressings are fixed at the value of the Landau gauge result at a perturbative scale.
We tested explicitly that the combination of these two procedures leads to cutoff independent results by varying $\Lambda^2$ by a factor of $10$.
\begin{table}[t]
 \begin{tabular}{|l|c||l|c|}
 \hline
  $\alpha_1^{\text{3g}}$ & $-\frac{9 \xi -17}{3 (\xi -3)}$ & $\alpha_1^{\text{gg}}$ & $-\frac{2\xi}{\xi-3}$ \\
  \hline
  $\alpha_2^{\text{3g}}$ & $-\frac{4 (3 \xi -2)}{3 (\xi -3)}$  & $\alpha_2^{\text{gg}}$ & $-\frac{2 (3 \xi +13)}{3 (\xi -3)}$ \\
  \hline 
  $\alpha_3^{\text{3g}}$ & $-\frac{2 (\xi +3)}{\xi -3}$ & & \\
 \hline
 \end{tabular}
\caption{\label{tab:exps_RGI}The exponents $\alpha$ for the RG improvement terms. The $\beta$'s are all $0$ because the exponents were derived as if the ghost dressing function were IR constant.}
\end{table}

Other ingredients for the numerical calculation are functions for extrapolating the dressings beyond the regime where they are calculated. For the UV the perturbative expressions
\begin{align}
 G^{\text{UV}}(x)=G(s)\left(\omega \ln\left(\frac{\LL}{\LQ}\right)\right)^\delta,\\
 Z^{\text{UV}}(x)=Z(s)\left(\omega \ln\left(\frac{\LL}{\LQ}\right)\right)^\gamma,
 \end{align}
are used. The anomalous dimensions can be found in Tab.~\ref{tab:anom_dims}. At low momenta we employ a simple extrapolation of the form $a\,p^2$ for the gluon propagator. For the ghost propagator we use a constant for $\xi=0$ and $a/(b+\ln p^2)$ for $\xi>0$.

\bibliographystyle{utphys_mod}
\bibliography{literature_linCovGauges_dses}

\end{document}